\documentclass[aps,prl,twocolumn,groupedaddress,showpacs]{revtex4}
\usepackage{graphicx}

\begin{document}


\title{Pressure-Tuned Collapse of the Mott-Like State in Ca$_{n+1}$Ru$_n$O$_{3n+1}$ (n=1,2):  Raman Spectroscopic Studies}



\author{C.S. Snow,$^1$ S. L. Cooper,$^1$
G. Cao,$^2,^3$ J.E. Crow,$^3$ H. Fukazawa,$^4$ S. Nakatsuji,$^3,^4$ Y. Maeno$^4$}

\affiliation{
$^{1}$Department of Physics and Frederick Seitz Materials Research Laboratory,
University of Illinois at Urbana-Champaign, Urbana, Illinois 61801\\
$^{2}$Department of Physics and Astronomy, University of Kentucky, Lexington, Kentucky  40506\\
$^{3}$National High Magnetic Field Laboratory, Tallahassee, Florida 32310\\
$^{4}$Department of Physics, Kyoto University, Kyoto 606-8502, Japan, and CREST, Japan Science and Technology Corporation, Japan}


\date{\today}

\begin{abstract}

We report a Raman scattering study of the pressure-induced collapse of the Mott-like phases of Ca$_3$Ru$_2$O$_7$ (T$_N$=56 K) and Ca$_2$RuO$_4$ (T$_N$=110 K).  The pressure-dependence of the  phonon and two-magnon excitations in these materials indicate: (i) a pressure-induced collapse of the antiferromagnetic (AF) insulating phase above P*$\sim$55 kbar in Ca$_3$Ru$_2$O$_7$ and P*$\sim$ 5-10 kbar in Ca$_2$RuO$_4$, reflecting the importance of Ru-O octahedral distortions in stabilizing the AF insulating phase; and (ii) evidence for persistent AF correlations above the critical pressure of Ca$_2$RuO$_4$, suggestive of phase separation involving AF insulator and ferromagnetic metal phases.

\end{abstract}


\maketitle


Complex oxide systems such as the cuprates, nickelates, and manganites have generated intense study over the past few years, in an effort, first, to clarify the relationship between the exotic phases of the materials; and second, to elucidate the origin and nature of the rich phenomena these compounds exhibit,\cite{1} such as unconventional superconductivity, charge- and orbital-ordering, and ``colossal'' magnetoresistance.  Although not so well studied, it is evident that the ruthenate system (Sr,Ca)$_{n+1}$Ru$_n$O$_{3n+1}$ ($n$= number of Ru-O layers per unit cell) exhibits the same richness and complexity as other complex oxides, including orbital-ordering in the Mott-like insulators Ca$_2$RuO$_4$ (T$_N$=110 K) and Ca$_3$Ru$_2$O$_7$ (T$_N$=56 K);\cite{2,3,4} ``colossal'' sensitivities of transport and structural properties to magnetic field,\cite{5} doping,\cite{6,7} and pressure;\cite{8} and a doping-induced transition in Ca$_{2-x}$Sr$_{x}$RuO$_4$ between a Mott insulator (x=0)\cite{6,7} and a possible unconventional superconductor with T$_C$=1.5 K (x=2).\cite{7,9,10}  A better understanding of these phenomena requires clarifying the manner in which the Mott-like phases of the ruthenates evolve into the exotic metallic and superconducting phases with doping or external perturbation. 

Measurements performed while pressure-tuning the phase behavior of a material are particularly valuable for carefully exploring the evolution of exotic phase behavior in complex materials:  such studies allow one to investigate - in a particularly well-controlled manner - the effects of doping on a system, absent the deleterious and obscuring effects of disorder introduced when doping via chemical substitution. 
For example, recent pressure-dependent transport and magnetization measurements of single-layer Ca$_2$RuO$_4$ have revealed a dramatic collapse of the antiferromagnetic (AF) insulator phase into a ferromagnetic (FM) metal phase near P*$\sim$5 kbar.\cite{8} Unfortunately, thus far there has been little information regarding the important evolution of lattice, charge, and spin {\it dynamics} across pressure-induced phase boundaries in these sytems.  In this Letter, we describe a novel investigation that combines the power of inelastic light (Raman) scattering for probing spin-, charge- and lattice-degrees of freedom in a material, with the systematic phase control afforded by pressure-tuning, in order to investigate the evolution of the spin-, charge- and lattice-dynamics through the pressure-tuned collapse of the Mott-like phases of Ca$_2$RuO$_4$ and Ca$_3$Ru$_2$O$_7$. This study allows us to investigate several critical issues related to these materials, including the role of structure and orbital polarization in stabilizing the AF insulating phases, and the manner in which AF correlations evolve into the pressure-induced metallic phases.

Raman scattering measurements were performed on single crystal samples of Ca$_2$RuO$_4$ and Ca$_3$Ru$_2$O$_7$ as functions of both temperature and hydrostatic pressure.  Single crystals were obtained by a floating zone technique for Ca$_2$RuO$_4$,\cite{7} and by a flux method for Ca$_3$Ru$_2$O$_7$.\cite{5}  The Raman spectra were taken in a true backscattering geometry with 647.1 nm incident photons from a Kr$^+$ laser.  Variable low-temperature high-pressure measurements were obtained by loading the samples into a SiC-anvil cell that was mounted on an insert specially-designed to fit in a flow-through helium cryostat, allowing continuous adjustment of both the temperature (3.5 - 300 K) and pressure (0 - 100 kbar).  Argon was used as the pressure-transmitting medium, and the pressure inside the cell was determined from the shift of the flourescence line of a small piece of ruby placed near the sample.  From the linewidth variation of the ruby line as a function of pressure and temperature, we estimate the degree of non-hydrostaticity in the pressure cell to be $\sim$1 \% or less.  The incident photon energy ($\sim$ 2 eV) is well below the $\sim$ 3 eV charge-transfer gap of the ruthenates;\cite{11} therefore pressure-dependent resonance enhancements of the scattering were not expected or observed for the excitations studied. 

We first focus attention on the Ca$_3$Ru$_2$O$_7$ system, which exhibits a transition from paramagnetic to AF states at T$_N$=56 K, and a metal-insulator transition at T$_{MI}$=48 K.\cite{12}  As first shown by Liu \emph{et al.},\cite{13} the metal-insulator transition in this material is clearly revealed in the Raman spectrum by the abrupt softening of the 438 cm$^{-1}$ out-of-phase (B$_{1g}$ symmetry) c-axis Ru-O phonon mode (see Fig. 1 (top)).  The softening of this mode is likely associated with an abrupt c-axis lattice contraction that is known to accompany the metal-insulator transitions of both Ca$_3$Ru$_2$O$_7$ \cite{5} and Ca$_2$RuO$_4$.\cite{6}  For our purposes, we use the phonon softening of this B$_{1g}$ mode to examine the pressure-dependence of the metal-insulator transition in Ca$_3$Ru$_2$O$_7$, T$_{MI}$(P).  Fig. 1 (middle,bottom) shows that the temperature at which the B$_{1g}$ phonon softening occurs systematically decreases with increasing pressure.  Identifying the mid-point temperature of this phonon softening as T$_{MI}$(P), we find a linear decrease of T$_{MI}$ with pressure - with a rate $\Delta T_{MI}/\Delta P \approx$ -0.85 K/kbar - and a complete collapse of the insulating phase above a critical pressure near P*$\sim$ 55 kbar (Fig. 1, bottom).

\begin{figure}
\centering
\includegraphics[width=6.75cm]{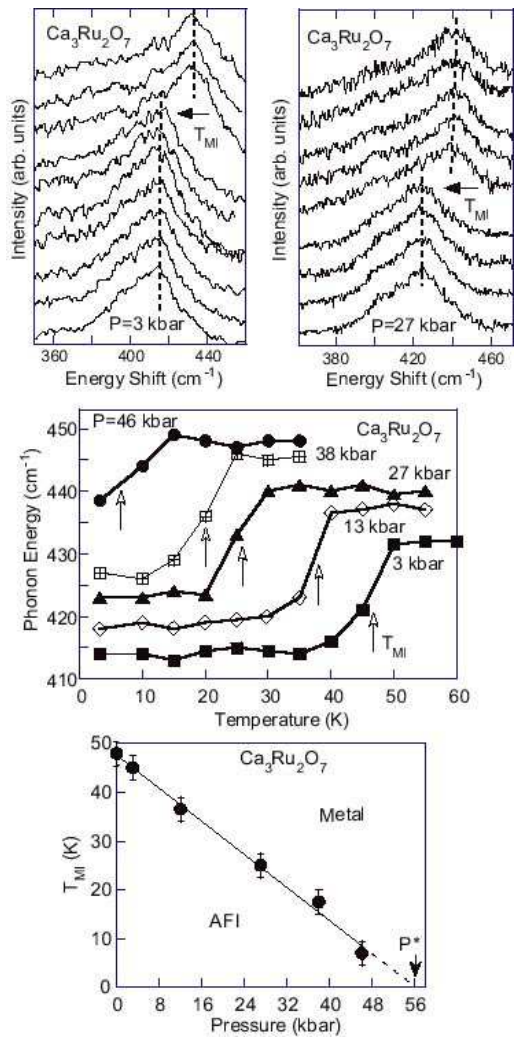}
\caption{Top:  P=3.5 kbar and 27 kbar Raman spectra of the B$_{1g}$ out-of-phase Ru-O phonon at temperatures (from bottom to top spectra): 3.5 K, 10 K, 15 K, 20 K, 30 K, 35 K, 40 K, 50 K, 55 K, and 60 K (3.5 kbar only).  The spectra have been offset.  Middle:  Summary of the temperature-dependence of the B$_{1g}$ out-of-phase Ru-O phonon at different values of the pressure.  The arrows denote the estimated metal-insulator transition, T$_{MI}$(P).  Bottom: Summary of the pressure-dependence of the metal-insulator transition temperature, T$_{MI}$(P).}\vspace{-0.4cm}
\end{figure}

The strong pressure dependence of T$_{MI}$ in Ca$_3$Ru$_2$O$_7$, and the eventual pressure-induced collapse of the insulating phase summarized in Fig. 1, clearly reveals behavior not described by the simple Mott-Hubbard description of the AF insulating phase.  In particular, this result suggests that the insulating phase of Ca$_3$Ru$_2$O$_7$ is stabilized by some structural parameter to which pressure strongly couples.  This parameter is most likely associated with Jahn-Teller-type distortions of the Ru-O octahedra, which are known to occur abruptly at the metal-insulator transitions of both Ca$_3$Ru$_2$O$_7$\cite{5} and Ca$_2$RuO$_4$.\cite{6}  A principal effect of these distortions is to reduce the orbital degeneracy of the Ru$^{4+}$ t$_{2g}$ levels by lowering the d$_{xy}$ orbital energy relative to those of the d$_{xz}$ and d$_{yz}$ orbitals.\cite{14}  Fang and Terakura have found in electronic structure calculations that these octahedral distortions stabilize an AF insulating phase by reducing the t$_{2g}$ bandwidth and shifting the Fermi surface nesting wavevector closer to the zone boundary.\cite{14}  Pressure, on the other hand, should thwart this tendency by reducing Ru-O octahedral distortions, thereby stabilizing the lower unit-cell volume metallic phase.\cite{15}

In view of the strong pressure-sensitivity of the Mott-like phase in the ruthenates, it is of interest to directly study how AF correlations evolve with pressure-tuning through the pressure-induced collapse of the AF insulating phase of Ca$_{n+1}$Ru$_n$O$_{3n+1}$.  We can directly probe this important issue by studying the pressure dependence of the two-magnon scattering response of Ca$_{n+1}$Ru$_n$O$_{3n+1}$.  Two-magnon scattering, which involves the simultaneous flip of spins on adjacent (Ru) sites, affords an ideal means of directly studying the short-range AF correlations in a system.\cite{13,16,17,18}  The associated Raman scattering process involves a photon-induced superexchange of two spins on nearest-neighbor (Ru 4d$^4$) sites; in the presence of AF order with an in-plane correlation length $\xi_{2D}$ greater than 2-3 lattice spacings, the energy cost of this spin-exchange process is 6.7J for a S=1 system.\cite{13,17,18}  Consequently, the energy, linewidth, and intensity of the two-magnon response afford substantial information regarding the exchange interaction energy and lifetime associated with AF fluctuations, even in the absence of long-range Ne\'{e}l order.

\begin{figure}
\centering
\includegraphics[width=6.75cm]{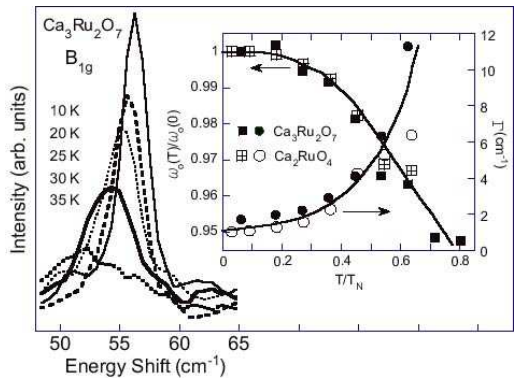}
\caption{P=0 two-magnon scattering response of Ca$_3$Ru$_2$O$_7$ at various temperatures.  Inset:  Summary of the normalized two-magnon energy, $\omega_o(T)/\omega_o(0)$, and linewidth, $\Gamma$, at P=0 as a function of the normalized temperature, T/T$_N$, for Ca$_2$RuO$_4$ [$\omega_o$(0)=102 cm$^{-1}$] (filled circle and square) and Ca$_3$Ru$_2$O$_7$ [$\omega_0$(0)=57 cm$^{-1}$] (open circle and square).  Lines drawn are guides to the eye.}\vspace{-0.4cm}
\end{figure}

Fig. 2 shows the two-magnon linewidth and normalized energy as a function of the normalized temperature, T/T$_N$, for Ca$_3$Ru$_2$O$_7$ and Ca$_2$RuO$_4$; illustrative two-magnon spectra are also shown at various temperatures for Ca$_3$Ru$_2$O$_7$ - the results observed for Ca$_2$RuO$_4$ are similar.  Identification of two-magnon scattering in Ca$_3$Ru$_2$O$_7$ and Ca$_2$RuO$_4$ is unambiguous, and is based upon the following features:  (a) the distinctive temperature dependence of this scattering in the Ne\'{e}l state (T$_N$ = 56 K in Ca$_3$Ru$_2$O$_7$; T$_N$ = 110 K in Ca$_2$RuO$_4$), namely, a characteristic decrease in energy and intensity, and an increase in linewidth, with increasing temperature, reflecting the renormalization of magnon energies and lifetimes by thermally-excited carriers;\cite{17,18} (b) the observed B$_{1g}$ symmetry of this scattering, which is that predicted by the two-magnon scattering Hamiltonian for a material with D$_{4h}$ symmetry,\cite{16} $\Sigma({\bf E}_I \cdot {\bf \sigma}_{ij})({\bf E}_S \cdot {\bf \sigma}_{ij}){\bf S}_i \cdot{\bf S}_j$, where {\bf E}$_I$ and {\bf E}$_S$ are the incident and scattered electric fields, ${\bf \sigma}_{ij}$ is a unit vector connecting sites i and j, and {\bf S}$_i$ is the spin operator on site i; and (c) the scaling of the two-magnon energies in Ca$_2$RuO$_4$  [$\hbar \omega_o$=102 cm$^{-1}$, T$_N$=110 K] and Ca$_3$Ru$_2$O$_7$ [$\hbar \omega_o$=57 cm$^{-1}$, T$_N$=56 K] with their respective Ne\'{e}l temperatures.  Using the energy cost of a two-magnon excitation in an S=1 layered antiferromagnet with z=4 nearest neighbors, $\hbar \omega_o$=6.7J, where J is the in-plane exchange energy, we estimate in-plane exchange energies of J=15.2 cm$^{-1}$ for Ca$_2$RuO$_4$ and J=8.5 cm$^{-1}$ for Ca$_3$Ru$_2$O$_7$.

The temperature-dependence of the two-magnon scattering responses in Ca$_2$RuO$_4$ and Ca$_3$Ru$_2$O$_7$ (inset, Fig. 2) is typical of that observed for AF insulators such as La$_2$NiO$_4$ \cite{18} and K$_2$NiF$_4$.\cite{17}  On the other hand, the pressure-dependence of the two-magnon response in the ruthenates, illustrated in Fig. 3 for Ca$_3$Ru$_2$O$_7$ [(a),(b)] and Ca$_2$RuO$_4$ [(c),(d)], is quite remarkable compared to that observed in more conventional Mott systems.  Specifically, the two-magnon energy in conventional Mott insulators such as La$_2$CuO$_4$\cite{19} and NiO\cite{20} systematically increases with increasing pressure:  $\Delta \omega_o/\Delta P \sim$ +2.6 cm$^{-1}$/kbar (+0.17\%/kbar) in NiO\cite{20} and $\Delta \omega_o/\Delta P \sim$ +4 cm$^{-1}$/kbar (+0.13\%/ kbar) in La$_2$CuO$_4$.\cite{19}  This pressure-dependence reflects an increase in the d-electron hopping matrix element t with increasing pressure, and a corresponding increase in the exchange constant J, in accordance with the simple Hubbard model, J=t$^2$/U, where U is the Coulomb interaction.  By contrast, Fig. 3 shows that the two-magnon energies in Ca$_3$Ru$_2$O$_7$ and Ca$_2$RuO$_4$ are relatively insensitive to pressure up to the pressure-induced insulator-metal transitions (e.g., $\Delta \omega_o/\Delta P$ $<$ -0.04\%/kbar in Ca$_3$Ru$_2$O$_7$), suggesting that the exchange interaction energy in the ruthenates is not appreciably influenced by pressure over this range.  Instead, increased pressure in the ruthenates primarily results in a systematic reduction in the intensity of two-magnon scattering.  These unusual trends in the pressure-dependence of the two-magnon scattering response suggest that in the ruthenates, increased pressure suppresses AF correlations, {\it not} via changes in the AF exchange coupling, but rather by suppressing the Ru-O octahedral distortions - and the resulting orbital polarization - that help stabilize antiferromagnetism in the ruthenates.\cite{2,5,6,14}

\begin{figure}
\centering
\includegraphics[width=9.25cm]{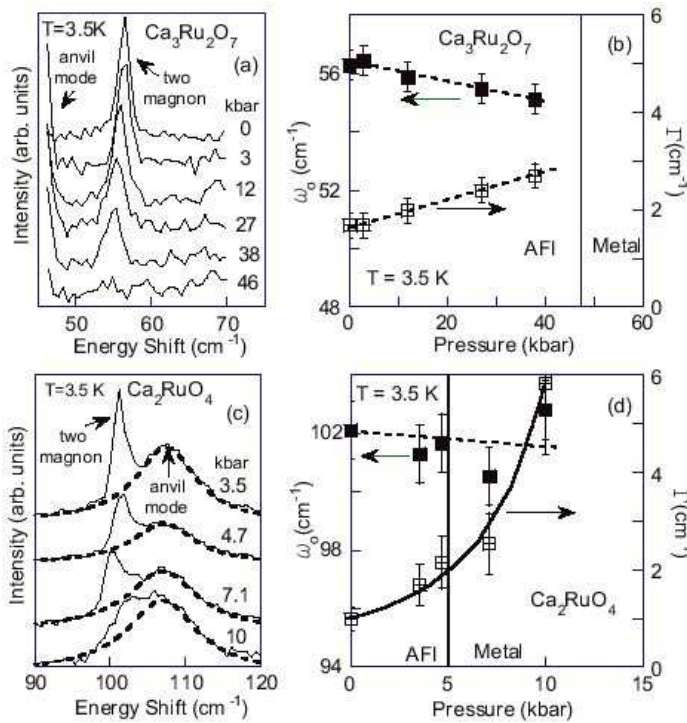}
\caption{Raman spectra of the two-magnon scattering response at different pressures for (a) Ca$_3$Ru$_2$O$_7$ and (c) Ca$_2$RuO$_4$.  The spectra have been offset.  Summary of the two-magnon energy $\omega$ and linewidth $\Gamma$ as a function of pressure for (b) Ca$_3$Ru$_2$O$_7$ and (d) Ca$_2$RuO$_4$.  Hatched lines reflect rough phase boundaries inferred from our phonon results for Ca$_3$Ru$_2$O$_7$, and from the results of reference~\cite{8} for Ca$_2$RuO$_4$.  Lines drawn are guides to the eye.  For Ca$_2$RuO$_4$, the two-magnon parameters were obtained after accounting for an anvil phonon mode with a simple lorentzian fit (dashed line).}\vspace{-0.4cm}
\end{figure}

A comparison of the pressure-dependences of Ca$_2$RuO$_4$ and Ca$_3$Ru$_2$O$_7$ is of particular interest (Fig. 3).  For example, although a pressure of P* $\sim$ 50 kbar is required to completely suppress AF correlations in Ca$_3$Ru$_2$O$_7$, AF correlations in Ca$_2$RuO$_4$ are completely suppressed at much lower pressures, P* $\sim$ 15 kbar, in spite of its substantially larger Ne\'{e}l temperature (T$_N$=110 K) and exchange coupling (J=15.2 cm$^{-1}$).  Moreover, while two-magnon excitations (and AF correlations) in Ca$_3$Ru$_2$O$_7$ are completely suppressed at the pressure-induced insulator-metal transition, in Ca$_2$RuO$_4$ we find that the two-magnon response persists well into the pressure-induced metallic phase.  Specifically, pressure-dependent transport and magnetization measurements of a Ca$_2$RuO$_4$ sample with a similar T$_N$  (=112 K) reveal a pressure-induced transition at P*=5 kbar from an AF insulator into a FM metallic phase;\cite{8} in support of this value for P*, we observe a large increase in the two-magnon linewidth of Ca$_2$RuO$_4$ above P=7 kbar, consistent with increased damping of magnons due to the presence of free carriers (Fig. 3 (d)).  However, our results clearly indicate that an observable two-magnon response persists to more than P $\sim$ 10 kbar, i.e., roughly twice the pressure at which the insulator-metal transition is induced in Ca$_2$RuO$_4$ at P*$\sim$5 kbar.  This indicates a possible phase separation regime involving FM metallic and AF insulating phases near P*, an interpretation that is consistent with the strongly first-order nature of the metal-insulator transition in these systems.

The dramatic sensitivity of T$_{MI}$ and AF correlations to pressure in the ruthenates, particularly Ca$_2$RuO$_4$,  is indicative of a strong competition between distinct orbital and magnetic phases in the ruthenates, with pressure-tuning pushing the system toward a pressure-tuned T=0 critical point at P* that separates the orbital-ordered AF insulating state and a pressure-induced FM metal phase.  Interestingly, calculations by Hotta and Dagotto\cite{3} and Anisimov et al.\cite{4} indicate the close proximity of AF and FM orbital-ordered phases, whose relative stability should be quite sensitive to pressure.  These calculations are consistent with our evidence for persistent AF correlations associated with a remnant AF insulating phase in the pressure-induced FM metal phase 
of Ca$_2$RuO$_4$.  Notably, theoretical estimates of the relative pressure-stability of different possible ordered phases in the ruthenates might help our results better differentiate among various orbital-ordering scenarios.

In summary, we have used a novel low-temperature, high-pressure Raman spectroscopic investigation to study the pressure-induced collapse of the AF insulating phases of Ca$_2$RuO$_4$ and Ca$_3$Ru$_2$O$_7$.  We find that, unlike conventional Mott insulators, pressure causes the AF insulating phase of the ruthenates to collapse at a T=0 critical point P*, most likely by suppressing Ru-O octahedral distortions - and the resulting d-orbital polarization - that stabilize the AF insulating phase of these 
systems.  Pressure-dependence studies of two-magnon scattering in Ca$_2$RuO$_4$ and Ca$_3$Ru$_2$O$_7$ reveal that the AF correlations in Ca$_2$RuO$_4$ are much more sensitive to pressure than in Ca$_3$Ru$_2$O$_7$, and that AF correlations associated with remnant AF insulator regions may persist well into the pressure-induced FM metallic phase of Ca$_2$RuO$_4$.  This is indicative of a strong competition between orbital-ordered AF insulating and FM metal phases, the outcome of which is strongly influenced by pressure.

We thank H. Rho for making some of the initial zero-pressure measurements on the Ca$_2$RuO$_4$ system, and we thank E. Dagotto, M. V. Klein, and F. Nakamura for useful discussions.  This work was supported by the Department of Energy under grant DEFG02-96ER45439.

\vspace{-0.7cm}

\bibliography{basename of .bib file}

\begin{thebibliography}{}

\vspace{-0.7cm}

\bibitem{1} M. Imada, A. Fujimori, and Y. Tokura, Rev. Mod. Phys. \textbf{70}, 1039 (1998).

\bibitem{2} S. Nakatsuji, S. Ikeda, and Y. Maeno, J. Phys. Soc. Jpn. \textbf{66}, 1868 (1997); G. Cao et al., Phys. Rev. B \textbf{56}, R2916 (1997); T. Mizokawa \emph{et al.}, Phys. Rev. Lett. \textbf{87}, 077202 (2001).

\bibitem{3} T. Hotta and E. Dagotto, Phys. Rev. Lett. \textbf{88}, 017201 (2001). 

\bibitem{4} V. I. Anisimov \emph{et al.}, Eur. Phys. J. B \textbf{25}, 191 (2002).

\bibitem{5} G. Cao \emph{et al.}, cond-mat/0205151 (2002).

\bibitem{6} M. Braden \emph{et al.}, Phys. Rev. B \textbf{58}, 847 (1998); O. Friedt \emph{et al.}, Phys. Rev. B \textbf{63}, 174432 (2001).

\bibitem{7} S. Nakatsuji and Y. Maeno, Phys. Rev. Lett. \textbf{84}, 2666 (2000); S. Nakatsuji and Y. Maeno, Phys. Rev. B \textbf{ 62}, 6458 (2000).

\bibitem{8} F. Nakamura \emph{et al.}, Phys. Rev. B \textbf{65}, 220402(R) (2002). 

\bibitem{9} T. M. Rice and M. Sigrist, J. Phys.:  Condensed Matter \textbf{7}, L643 (1995).

\bibitem{10} Y. Maeno \emph{et al.}, Nature (London) \textbf{372}, 532 (1994); A. P. Mackenzie \emph{et al.}, Phys. Rev. Lett. \textbf{80}, 161 (1998).

\bibitem{11} A. V. Puchkov \emph{et al.}, Phys. Rev. Lett. \textbf{81}, 2747 (1998).

\bibitem{12} G. Cao \emph{et al.}, Phys. Rev. B \textbf{56}, 5387 (1997).

\bibitem{13} H. L. Liu \emph{et al.}, Phys. Rev. B \textbf{60}, 6980 (1999).

\bibitem{14} Z. Fang and K. Terakura, Phys. Rev. B \textbf{64}, 020509(R) (2001); see also I.I. Mazin and D.J. Singh, Phys. Rev. Lett \textbf{82}, 4324 (1999).

\bibitem{15} A similar suppression of Jahn-Teller distortions with pressure was also observed in room-temperature Raman measurements of La$_{0.75}$Ca$_{0.25}$MnO$_4$; A. Congeduti et al., Phys. Rev. Lett. \textbf{86}, 1251 (2001).

\bibitem{16} G. Blumberg \emph{et al.}, Science \textbf{278}, 1427 (1997); M. R\"{u}bhausen \emph{et al.}, Phys. Rev. Lett. \textbf{82}, 5349 (1999).

\bibitem{17}  P. A. Fleury and H. J. Guggenheim, Phys. Rev. Lett. \textbf{24}, 1346 (1970); P. A. Fleury and R. Loudon, Phys. Rev. \textbf{166}, 514 (1968).

\bibitem{18} S. Sugai \emph{et al.}, Phys. Rev. B \textbf{42}, 1045 (1990); S. Sugai \emph{et al.}, J. Phys. Soc. Japan \textbf{67}, 2992 (1998).

\bibitem{19} M. C. Aronson \emph{et al.}, Phys. Rev. B \textbf{44}, 4657 (1991). 

\bibitem{20} M. J. Massey \emph{et al.}, Phys. Rev. B \textbf{42}, 8776 (1990).
\end{thebibliography}

\end{document}